\documentclass[useAMS,usenatbib]{mn2e}
\usepackage{natbib}
\usepackage{graphicx}
\newcommand{\eqb}{\begin{equation}}
\newcommand{\eqe}{\end{equation}}
\begin{document}

\title{Transformation of the Poynting flux into the kinetic energy in relativistic jets}
\author[Y.E.Lyubarsky]{Y.E.Lyubarsky\\
Physics Department, Ben-Gurion University, P.O.B. 653, Beer-Sheva
84105, Israel; e-mail: lyub@bgumail.bgu.ac.il}
\date{Received/Accepted}
\maketitle
\begin{abstract}
The acceleration of relativistic jets from the Poynting to the matter dominated stage is considered. The are generally two collimation regimes, which we call equilibrium and non-equilibrium, correspondingly. In the first regime, the jet is efficiently accelerated till the equipartition between the kinetic and electro-magnetic energy. We show that after the equilibrium jet ceases to be Poynting dominated,
the ratio of the electro-magnetic to the kinetic energy decreases only logarithmically
so that such jets become truly matter dominated only at extremely large distances.  Non-equilibrium jets remain generally Poynting dominated till the logarithmically large distances. In the only case when a non-equilibrium jet is accelerated till the equipartition level, we found that the flow is not continued to the infinity but is focused towards the axis at a finite distance from the origin.
\end{abstract}
\begin{keywords}
 MHD -- galaxies:jets -- gamma-rays:bursts
\end{keywords}

\section{Introduction}

Collimated, Poynting dominated outflows are considered as a viable model for relativistic jets in active galactic nuclei (AGNs), microquasars and gamma-ray bursts (GRBs).
For the relativistic outflows, the question of how the electromagnetic energy is transformed into the plasma energy has no simple answer.
%The central issue of the theory of Poynting dominated outflows is the transformation of .
%This problem is specific for relativistic flows.
In the non-relativistic case, the Poynting flux is efficiently converted into the kinetic energy of the flow; an approximate equipartition is reached already at the Alfven point, where the toroidal
magnetic field becomes comparable with the poloidal field. Relativistic flows remain Poynting dominated even at the fast magnetosonic point. The reason is that in this case, the magnetic force is balanced not by inertia but by the electric force so that the plasma is only weakly accelerated by a small residual of the magnetic and electric forces. It turns out that in unconfined, nondissipative flows,
%Most of the electro-magnetic energy could be converted to the plasma energy only in the far zone however,
the characteristic energy transformation scale is inadmissibly large; such a flow is accelerated only to the Lorentz factor of the order of $\gamma_{\rm max}^{1/3}$, where $\gamma_{\rm max}$ is the Lorentz factor corresponding to the total transformation
of the electro-magnetic into the kinetic energy, after which acceleration practically ceases,
$\gamma\sim(\gamma_{\rm max}\ln r)^{1/3}$ (Tomimatsu 1994; Beskin, Kuznetsova \& Rafikov 1998).

The electro-magnetic energy could be more efficiently converted into the
kinetic energy if the flow is
collimated by an external medium. Such a configuration arises naturally in gamma-ray bursts,
where the relativistic jet from the collapsing stellar core pushes its way through the stellar envelope.
In the accreting systems, the magnetically driven outflow from the rotating black hole could be collimated by the pressure of a slow
(and generally magnetized) wind from the outer parts of the accretion disk.
Collimation and acceleration of externally confined, Poynting dominated jets has
being studied extensively both numerically and analytically, see the resent works by
Komissarov et al. (2007, 2009), Narayan et al. (2008), Lyubarsky (2009, thereafter Paper I)
and references therein.

In the simplest case of the power-law external pressure distribution, $p\propto z^{-\kappa}$, the conditions for the flow acceleration and collimation are the following (Paper I):
\begin{enumerate}
\item at $\kappa>2$, the flow becomes asymptotically radial and the acceleration is practically saturated at
$\gamma\sim (\gamma_{\rm max}/\Theta^2)^{1/3}$, where $\Theta$ is the final collimation angle,
which itself is determined by the outer pressure distribution;
 \item at $\kappa\le 2$, the flow is accelerated until it ceases to be Poynting dominated; the shape of the flow line is paraboloidal, $r\sim z^k$,
   where $k<1$ is determined by the outer pressure distribution.
\end{enumerate}
One sees that in the scope of ideal MHD, the electro-magnetic energy is efficiently converted into the kinetic energy
only if the flow is confined by the external medium with the pressure decreasing not too fast.
However, one has to stress that at $\kappa<2$, the flow Lorenz factor grows proportionally
to the jet radius so that if the pressure decreases too slowly, so that the flow expands slowly, the acceleration rate would also be very low.
%Namely, the plasma is accelerated only in expanding flows, which assumes that the surrounding pressure decreases.
In particular, if the surrounding pressure goes to a constant,
the flow becomes cylindrical and stops accelerating.

Till now only acceleration of the Poynting dominated flows has been addressed; the
results obtained were just extrapolated to the energy equipartition stage.
The transition from the Poynting dominated to the matter dominated stage has not been
studied yet. An important point is
that the flow could be considered as truly matter dominated if the ratio of the Poynting
to the kinetic energy flux, $\sigma$, becomes less than approximately 0.1. The reason is
that only in this case, the shock jump conditions become close to those
in the unmagnetized medium %and a significant fraction of the upstream energy flux
%could be converted into the thermal energy
(Kennel \& Coroniti 1984; Appl \& Camenzind 1988) so that the interaction of the
jet with the surroundings occurs as in the non-magnetized case.
At $\sigma>0.1$, only weak shocks are possible
therefore the flow pattern, which arises when such a  jet
is decelerated in the ambient medium, significantly differs from that for the
purely hydrodynamic jet (Komissarov 1999). Simulations show that already at $\sigma\approx 0.01$, the flow pattern differs significantly from the purely hydrodynamic one
(Leismann et al 2005). In the GRB context, Mimica, Giannios and Aloy (2009) and  Mimica and Aloy (2009) show that even a moderate magnetization
of the ejecta could have a profound effect on the properties of the internal shocks as well as on
dynamics of the deceleration thus affecting both the prompt and the afterglow emission.

Here we study the transition of the flow through the $\sigma\sim 1$ domain.
We address only the case $\kappa\leq 2$ when such a transition could occur at all.
The paper is organized as follows.
In Section 2, we present the asymptotic equations describing relativistic, magnetized flows
at large distance from the origin. In Section 3, we find solutions to these equations in the case $\kappa<2$. The case $\kappa=2$ is addressed in
Section 3. Conclusions are presented in Section 5.

\section{The jet structure in the far zone}

In Paper I, we presented asymptotic equations describing the relativistic, magnetized flow
at distances much larger than the light cylinder radius.
Now we shortly outline the relevant results.
As usual, the magnetic field is
conveniently decomposed into the poloidal and toroidal
components,
$\mathbf{B}=\mathbf{B}_p+B_{\phi}\mathbf{\widehat{\phi}}$, the
poloidal field being expressed via the flux function
 \eqb
 \mathbf{B}_p=\frac 1r\nabla\Psi\times\mathbf{\widehat{\phi}}.
 \label{Bfield}\eqe
We use cylindrical $(r,\phi,z)$ coordinates; the hat denotes
unit vectors.   The distribution of the
mass flux at the inlet of the flow is described by the function $\eta(\Psi)$
defined by the continuity equation
 \eqb
\eta(\Psi)=\frac{4\pi\rho v_p\gamma}{B_p};
 \label{continuity}\eqe
where $\rho$ is the plasma density in the lab frame, $\gamma$ the flow Lorentz factor, $v_p$ the poloidal velocity.
The conserved total energy flux is presented as
 \eqb
 \gamma-\frac{r\Omega B_{\phi}}{\eta}=\mu(\Psi);
 \label{energy}\eqe
where $\Omega(\Psi)$ is the angular velocity of the flied line.
In this expression, the first term is the kinetic energy whereas the second one
is the Poynting flux.

The energy integral $\mu$ is determined from the
condition of smooth passage of the flow through the singular points.
In the far zone, it should be considered as a given function.
An important point is that close to the axis,
the energy integral has the universal form
 \eqb
\mu(\Psi)=\gamma_{\rm
in}\left(1+\frac{\Psi}{\widetilde{\Psi}}\right);\quad
\widetilde{\Psi}=\frac{\gamma_{\rm in}\eta}{2\Omega^2};
 \label{energy1}\eqe
where $\gamma_{\rm in}$ is the Lorentz factor at the inlet of the flow.
In this expression, the first and the second term describe the kinetic and
the Poynting energy flux,
correspondingly, at the inlet of the flow. Note that the Poynting flux goes to zero at the axis therefore the flow is Poynting dominated only at $\Psi\gg\widetilde{\Psi}$.
%Note that the widely used parameter
%$\sigma$, defined as the ratio of the Poynting to the matter
%energy flux, is presented via the basic quantities as
% \eqb
%\sigma=\frac{\mu-\gamma}{\gamma}.
% \eqe

The structure of the flow is described by the transfield and Bernoulli equations.
As an unknown function, one can conveniently use the shape of the
magnetic flux surface, $r(z,\Psi)$. For collimated flows, $r\ll z$, the
transfield force balance equation in the far
zone could be written as
\begin{eqnarray}
\eta\mu\left[-\frac{\partial^2r}{\partial
z^2}+\frac{1}{\Omega^2r^3}\left(1-\frac{2\gamma_{\rm
in}}{\mu}+\frac{\gamma_{\rm in}^2}{\gamma^2}\right) \right] \nonumber \\
=\frac1{2r}\left(1+\frac{\gamma_{\rm in}^2}{\Omega^2r^2}\right)
\frac{\partial}{\partial\Psi}\frac{\eta^2(\mu-\gamma)^2}{\Omega^2\gamma^2}.
\label{asympt_transfield}\end{eqnarray}
The Bernoulli equation is reduced, beyond the fast magnetosonic point, to
 \eqb
\eta(\mu-\gamma)\frac{\partial r}{\partial\Psi}=\Omega^2r.
 \label{Bernoulli0}\eqe
%The set of equations (\ref{asympt_transfield}) and (\ref{Bernoulli0}) describes the %relativistic, magnetized jets fa away from the origin.

Here we are interested in outflows subtending a finite magnetic flux $\Psi_0$ therefore Eqs.
(\ref{asympt_transfield}) and (\ref{Bernoulli0}) should be solved at $0\le\Psi\le\Psi_0$.
We assume that $\Psi_0\gg\widetilde{\Psi}$ so that the main body of the flow is
initially Poynting dominated.
If the flow is confined by the pressure of the external medium, $p_{\rm ext}(z)$, the pressure
balance condition should be satisfied at the boundary:
 \eqb
\left(\frac{\eta(\mu-\gamma)}{\Omega
r\gamma}\right)^2_{\Psi=\Psi_0}=8\pi p_{\rm ext}(z).
 \label{boundary}\eqe
The boundary condition at the axis is $r(\Psi=0)=0$.

%Transition to $\sigma\sim 1$ is described by the asymptotic
%transfield equation (\ref{Poynting_jet}) complemented by the
%Bernoulli equation in the form (\ref{Bernoulli0}). Generally
%this could be done only numerically. Here we analyze the
%equilibrium collimation when the solution could be obtained semi-analytically. when one can neglect the %curvature term
%in the transfield equation. Then the transfield equation
%becomes an ordinary differential equation and the jet structure
%could be found easily in the same way as the structure of the
%core was found in the previous section.

%The equilibrium core is present close enough to the axis of any jet; the structure of such a core
%was studied in Paper I. It was also shown that the whole jet is collimated in the equilibrium regime
%if the external pressure decreases slower than $z^{-2}$. Then the flow is collimated and accelerated
%until it ceases to be Poynting dominated. In this paper, we study the structure of the equilibrium jet
%relaxing the assumption that the main body of the flow is Poynting dominated. We will show that
%even when the equilibrium jet ceases to be Poynting dominated, the flow is collimated further out
%approaching cylindrical shape at infinity. The Poynting flux continuously decreases down to zero.

There are generally two different regimes of collimation. At the condition
 \eqb
r\left\vert\frac{\partial^2r}{\partial
z^2}\right\vert\ll \frac 1{\Omega^2r^2},
  \label{equilibrium}\eqe
one can neglect the term with the derivative in $z$ and write
the transfield equation (\ref{asympt_transfield}) as an ordinary differential equation
 \eqb
\mu\left(1+\frac{\gamma_{\rm
in}^2}{\gamma^2}\right)-2\gamma_{\rm in}=
\frac{\Omega^2r^2+\gamma_{\rm in}^2}{\Omega\gamma}(\mu-\gamma)
\frac{\partial}{\partial\Psi}\frac{\eta(\mu-\gamma)}{\Omega\gamma}.
 \label{axis_transfield}\eqe
This equation describes in fact cylindrical equilibrium, in which case the residual of the
magnetic hoop stress and the electric force is counterbalanced by the pressure of
the poloidal field. The corresponding collimation regime is called equilibrium because in this case, the structure of the jet at any distance from the origin is the same
as the structure of an appropriate equilibrium cylindrical configuration. For a smoothly expanding jet,
the condition (\ref{equilibrium}) is reduced to
\eqb
 \Omega r^2\ll z;
 \label{equilibrium_cond}\eqe
Note that neglecting the second derivative in the transfield equation (\ref{asympt_transfield}),
one looses solutions. These lost solutions just describe oscillations of the flow with respect to the equilibrium state satisfying Eq. (\ref{axis_transfield}). Therefore if the condition (\ref{equilibrium_cond})
is fulfilled, one can anyway use Eq. (\ref{axis_transfield}) in order to find the overall expansion of the jet.

The transverse equilibrium implies that in the proper plasma frame, the toroidal and poloidal magnetic fields are comparable, $B'_p\sim B'_{\phi}$. Transforming to the lab frame, one gets $\gamma\sim B_{\phi}/B_p$. Taking into account that the toroidal field is wound up from the poloidal one so that
$B_{\phi}\approx\Omega rB_p$, one concludes that in the equilibrium flow, $\gamma\sim\Omega r$. This estimate is confirmed by explicit solutions (Tchekhovskoy et al 2008; Komissarov et al 2009; Paper I;
Beskin \& Nokhrina 2009). Now one sees that the condition (\ref{equilibrium_cond}) in fact implies that the flow is in causal connection, i.e. the proper propagation time $z/\gamma$ exceeds the time $r$ necessary for a signal to cross the flow so that the flow has enough time in order to settle into transverse equilibrium.

If the condition opposite to (\ref{equilibrium_cond}) is
fulfilled,
\eqb
 \Omega r^2\gg z;
 \label{nonequilibrium_cond}\eqe
which anyway could happen only far enough from
the axis, the term with the second derivative becomes
dominant and the transfield equation is reduced to
 \eqb
-2\mu\eta r\frac{\partial^2r}{\partial z^2}=
\frac{\partial}{\partial\Psi}\frac{\eta^2(\mu-\gamma)^2}{\Omega^2\gamma^2}.
 \label{nonequilibrium}\eqe
This equation could be directly obtained neglecting the poloidal magnetic field and
the azimuthal velocity (Lyubarsky \& Eichler 2001).
Such a flow may be conceived as composed from coaxial magnetic loops.
We call the corresponding collimation regime non-equilibrium. Non-equilibrium flows are causally disconnected or marginally connected therefore the magnetic loops generally do not shrink even though the poloidal field pressure is negligibly small (see, however, sect 4).

In Paper I, the structure of the jet was found under the condition that the main body of the flow
remains Poynting dominated. Here we relax this condition and study the transition to the matter dominated flow.
We consider the jet with a constant angular velocity, $\Omega=\it const$ and homogeneous injection, $\eta=\it const$; $\gamma_{\rm in}=\it const$. In this case, one can conveniently use the dimensionless
variables
 \eqb
X=\Omega r;\quad Z=\Omega z.
 \eqe
We also assume
that the energy integral, $\mu(\Psi)$, is described by a linear function (\ref{energy1}) throughout the flow. The last is
a good approximation for jets with a constant angular velocity (Komissarov et al 2007, 2009; Tchekhovskoy
et al 2008).
We assume that the external pressure decreases as a power law
 \eqb
p=\frac{p_0}{Z^{\kappa}};\quad p_0=\frac{\beta\Omega^4\Psi_0^2}{6\pi}.
 \label{pressure}\eqe
The normalization coefficient $\beta$ is chosen as in Paper I.

\section{Equilibrium jets}

The Poynting dominated jet is collimated in
the equilibrium regime if $\kappa<2$. Then the jet expands as
$X\propto Z^{\kappa/4}$ whereas the Lorentz
factor increases as $\gamma\sim X$ (Tchekhovskoy et al 2008; Komissarov et al 2009; Paper I;
Beskin \& Nokhrina 2009).
Now we address the transition of the equilibrium jet from the Poynting dominated
to the kinetic energy dominated flow.

In the equilibrium regime, the flow is described by a pair of ordinary differential equations
(\ref{Bernoulli0}) and (\ref{axis_transfield}) for the transverse
structure of the jet, the dependence on $Z$ entering only via the boundary condition (\ref{boundary}). Introducing the variables
 \eqb
\xi=\frac X{\gamma_{\rm in}},\quad
\Gamma=\frac{\gamma}{\gamma_{\rm in}},\quad
s=1+\frac{\Psi}{\widetilde{\Psi}},
  \eqe
one reduces Eqs. (\ref{Bernoulli0}) and (\ref{axis_transfield})
to the dimensionless form
  \begin{eqnarray}
\frac s{\Gamma}\frac{d\Gamma}{ds} & = & 1-\frac{s+\Gamma^2(s-2)}{2(1+\xi^2)(s-\Gamma)};\label{core_tr1}\\
\frac{d\xi}{ds} & = & \frac{\xi}{2(s-\Gamma)}.\label{core_Bern1}
  \end{eqnarray}
Solutions to these equations were analyzed in detail in Paper I.
Any solution describes the
transverse structure of the jet at some distance from the origin.
Near the axis, $\xi\ll 1$, the solution is
 \eqb
 s=1+C\xi^2;\qquad \Gamma=1+\frac 12
C\xi^4;
 \label{init1}\eqe
where $C$ is a constant. If $C>0.38$, the solution goes to a Poynting dominated flow
far from the axis, $\xi\gg 1$. In the opposite case, $C<0.38$, the solution
describes the transverse structure of the matter dominated jet.

For any constant $C$, the solution to Eqs. (\ref{core_tr1}) and (\ref{core_Bern1}) could be easily found numerically by the Runge-Kutta method. In order to find the full structure of the jet,
one has to find an appropriate constant $C$ for any distance from the origin, $Z$.
 %describing the transverse structure of the jet at this distance.
We are looking for a solution satisfying the outer boundary condition (\ref{boundary}),
which is written in the new variables as
 \eqb
\left.\frac{s-\Gamma}{\xi\Gamma}\right\vert_{s=1+\Psi_0/\widetilde{\Psi}}=
\sqrt{\frac{\beta}3}\frac{\gamma_{\rm in}\gamma_{\rm max}}{Z^{\kappa/2}};
 \label{boundary1}\eqe
where
 \eqb
\gamma_{\rm max}=\mu(\Psi_0)=\gamma_{\rm in}\frac{\Psi_0}{\widetilde{\Psi}}
 \label{gamma_max}\eqe
is the maximal achievable Lorentz factor of the flow, which is just Michel's
magnetization parameter (Michel 1969).
At any $Z$, one has to find a constant $C$ such that the solution to Eqs. (\ref{core_tr1}) and (\ref{core_Bern1}) satisfies both the condition (\ref{init1}) and the condition (\ref{boundary1}).
This could be easily done by bisection.

As an example, we presented in Fig. 1 the structure of the jet confined by the pressure $p\propto z^{3/2}$. One sees that initially  the flow expands according to the Poynting dominated scaling $X\propto Z^{\kappa/4}=Z^{3/8}$. A cylindrical,
moderately magnetized, $\sigma\sim 1$, core is formed within the jet at this stage, as it was discussed in Paper I. When the bulk
of the flow ceases to be Poynting dominated, the jet begins to expand faster than at the Poynting dominated stage. Note that in the logarithmic plot, it looks as if the jet inflates however, the jet collimation angle, $X/Z$, still decreases but slower than at the Poynting dominated stage. The weaker collimation of the flow in the matter dominated stage has already been noticed by Komissarov et al (2009).
In Fig. 2, we show evolution
of the Lorentz factor and of the magnetization parameter along selected flow lines.
Note that according to the general equilibrium scaling $\gamma\sim X$, the Lorentz factor
 is larger at the periphery of the flow. One sees also that even though $\sigma$ continuously decreases with the distance, the $\sigma$ decreasing rate drops down after the flow reaches equipartition therefore the flow becomes truly
matter dominated ($\sigma<0.1$) only at extremely large distances.

\begin{figure*}
\includegraphics[scale=0.7]{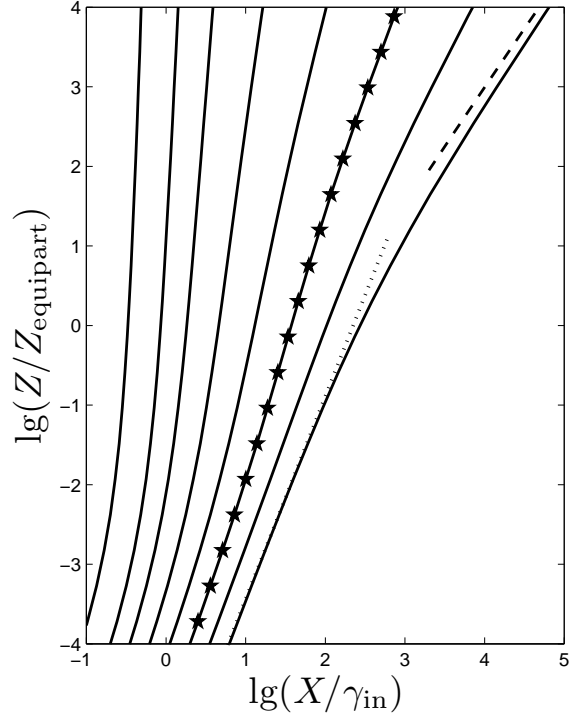}
\caption{The flux surfaces in the equilibrium jet;  $\kappa=1.5$;
$\gamma_{\rm max}/\gamma_{\rm in}=200$. The equipartition radius is defined by Eq. (\ref{Ztr}). Dotted line shows the Poynting dominated asymptotics
(\ref{radius}) whereas dashed line shows the matter dominated asymptotics (\ref{shape}).
The discrepancy between the last and the numerical solution is attributed to the fact that
even at the distances as large as $10^4Z_{\rm equipart}$, $\sigma$  is still
not very small (see Fig. 2).}
\end{figure*}
\begin{figure*}
\includegraphics[scale=0.7]{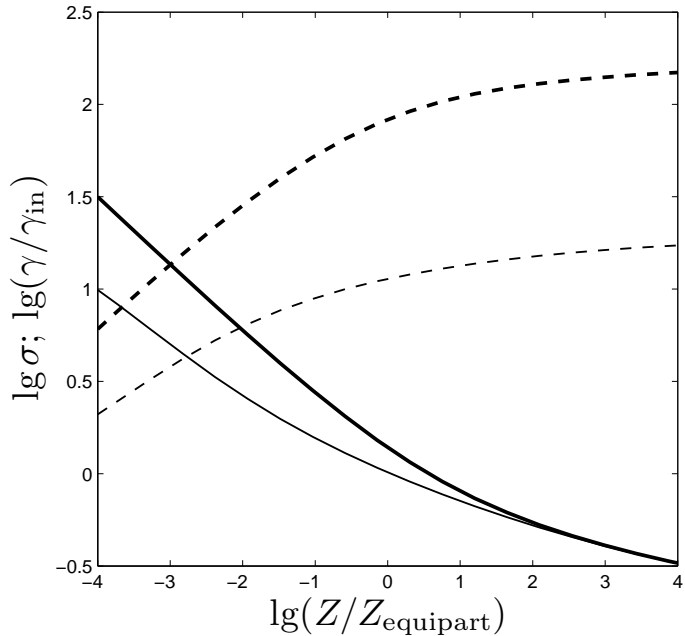}
\caption{Evolution of the Lorentz factor (dashed) and of the
ratio of the Poynting to the kinetic energy flux (solid) along
the boundary of the jet (thick lines) and along
the flow line marked by asterisks in Fig. 1 (thin lines). One sees that, according to the general
equilibrium scaling $\gamma\sim \Omega r$, the Lorentz factor is larger at the periphery than within the jet. One also sees that at $Z\gg Z_{\rm equipart}$, $\sigma$ becomes independent of $r$ and decreases with $z$ very slowly; this is confirmed by the asymptotic result of Eq. (\ref{sigma_small}).}
\end{figure*}

Let us now find analytically asymptotic solutions describing equilibrium flows in the
Poynting and matter dominated domains, correspondingly.
With this purpose, one can conveniently use $\sigma$ as a variable instead of $s$:
\eqb
\sigma=\frac s{\Gamma}-1.
  \label{sigma}
\eqe
Then Eqs. (\ref{core_tr1}) and (\ref{core_Bern1}) are reduced to
  \begin{eqnarray}
&&\xi(1+\xi^2)\frac{d\sigma}{d\xi}=(\Gamma-1)^2+\sigma(1+\Gamma^2);\label{equil1}\\
&&\frac{(1+\sigma)\xi(1+\xi^2)}{\Gamma}\frac{d\Gamma}{d\xi}=(1+2\xi^2-\Gamma^2)\sigma-(\Gamma-1)^2.
 \label{equil2}  \end{eqnarray}
The condition at the axis is $\sigma(0)=0$. The outer boundary of the jet, $\xi_0$, is defined
by the condition
\eqb
s(\xi_0)=1+\Psi_0/\widetilde{\Psi}\approx\Psi_0/\widetilde{\Psi}=\gamma_{\rm max}/\gamma_{\rm in};
 \eqe
then Eq. (\ref{sigma}) yields
 \eqb
\sigma(\xi_0)=\frac{\gamma_{\rm max}}{\gamma_{\rm in}\Gamma(\xi_0)}-1.
 \label{boundary3}\eqe
The boundary condition (\ref{boundary1}) is written in the new variables as
 \eqb
\frac{\sigma(\xi_0)}{\xi_0}=
\sqrt{\frac{\beta}3}\frac{\gamma_{\rm in}\gamma_{\rm max}}{Z^{\kappa/2}}.
 \label{boundary2}\eqe

Let us first find the solution for the Poynting dominated part of the jet, $\sigma\gg 1$.
Note that since the Poynting flux goes to zero at the axis of the flow (see Eq. (\ref{energy1})),
the condition $\sigma\gg 1$ could be met only at $\xi\gg 1$.
Moreover, we can take $\Gamma\gg 1$
in this range because the Poynting dominated flow is accelerated so that eventually the Lorentz factor of the flow exceeds the initial one. Then Eqs. (\ref{equil1})
and (\ref{equil2}) are reduced to
 \begin{eqnarray}
 && \xi^3\frac{d\sigma}{d\xi}=\sigma\Gamma^2;\label{eq1}\\
  &&\frac{\xi^3}{\Gamma}\frac{d\Gamma}{d\xi}=2\xi^2-\Gamma^2.\label{eq2}
 \end{eqnarray}
The solution to the second equation is
 \eqb
\Gamma=\frac{\xi}{\sqrt{1+\exp(-D\xi^2)}};
 \eqe
where $D$ is a constant.
Since Eqs. (\ref{eq1}) and (\ref{eq2}) are valid only far from the axis,
one has to solve them in the limit $\xi\gg 1$; this yields
 \eqb
\Gamma=\xi;\quad \sigma=A\xi;
 \label{gamma_sigma}\eqe
where $A$ is a constant, which is found from the boundary condition (\ref{boundary2}) as
 \eqb
A= \sqrt{\frac{\beta}3}\frac{\gamma_{\rm in}\gamma_{\rm max}}{Z^{\kappa/2}}.
 \label{A}\eqe
Note that the $\xi\gg 1$ solution is independent of the constant $D$ therefore the structure of the flow far from the axis is uniquely determined by the outer boundary condition (\ref{boundary2}). The condition at the axis of the flow, Eq. (\ref{init1}), does not place any restriction on the structure of the Poynting dominated flow at $\xi\gg 1$.

The expressions (\ref{gamma_sigma}) and (\ref{A}) describe the internal structure of
the equilibrium jet not too close to the axis. One sees that the Lorentz factor at any point of the flow
is proportional to the cylindrical radius of the point,
 \eqb
\gamma=X;
\label{gamma_eq}\eqe
so that in any cross-section of the jet the Lorentz factor increases outwards whereas at any flux surface it increases with the distance so far as the flow expands. This is the general property of equilibrium
Poynting dominated jets (Tschekovskoy et al 2008, Paper I, Beskin \& Nokhrina 2009). The jet radius may be found by substituting Eqs. (\ref{gamma_sigma}) to the condition (\ref{boundary3}) and taking into account that $\sigma\gg 1$; this yields
 \eqb
X_0=\gamma_{\rm in}\xi_0=\left(\frac{3Z^{\kappa}}{\beta}\right)^{1/4}.
 \label{radius}\eqe
This expression was already obtained, by different methods, in
(Tschekovskoy et al 2008, Komissarov et al 2009, Paper I).

The presented solution remains valid while $\sigma$ remains large, i.e. outside the boundary
defined from Eqs. (\ref{gamma_sigma}) and (\ref{A})
 \eqb
X_{\rm core}=\sqrt{\frac 3{\beta}}\frac 1{\gamma_{\rm max}}Z^{\kappa/2}.
 \label{core}\eqe
Near the axis, a $\sigma\approx 1$ core is formed (Paper I). Comparing Eq. (\ref{radius})
with Eq. (\ref{core}), one sees that the fraction of the jet volume occupied by the core
grows with the distance from the origin therefore
eventually the whole jet ceases to be Poynting dominated.
%. According to Eq. (\ref{gamma_sigma}), this happens when $\xi_0=\gamma_{\rm max}/\gamma_{\rm in}$. Then Eq. (\ref{gamma_sigma}) yields the characteristic
%equipartition distance from the
This happens at the distance
 \eqb
Z_{\rm equipart}=\left(\frac{\beta\gamma^4_{\rm max}}3\right)^{1/\kappa}.
 \label{Ztr}\eqe
The full solution presented in Figs. 1 and 2 confirms this scaling.

Let us now solve Eqs. (\ref{equil1})
and (\ref{equil2}) in the zone $Z\gg Z_{\rm equipart}$,
where the Poynting flux is already small as compared with the kinetic energy flux,
$\sigma\ll 1$.
Inspecting Eq. (\ref{equil1}), one sees that in order for $\sigma$
to be small, $\Gamma$ should be close to unity at $\xi\sim 1$.
Moreover, comparing the left-hand side of Eq. (\ref{equil2}) with the first term in the right-hand side, one sees that
there should be $\Gamma-1\sim\sigma$. Then one can neglect $(\Gamma-1)^2$ with respect to $\sigma$
and reduce Eqs. (\ref{equil1}) and (\ref{equil2}) to
  \begin{eqnarray}
&&\xi(1+\xi^2)\frac{d\sigma}{d\xi}=2\sigma;\label{22}\\
&&\frac{\xi(1+\xi^2)}{\Gamma}\frac{d\Gamma}{d\xi}=2\sigma\xi^2.\label{23}
  \end{eqnarray}
The solution to Eq. (\ref{22}) is
 \eqb
\sigma=\frac{C\xi^2}{1+\xi^2};
\label{small_sigma1}\eqe
where $C$ is a constant.

One sees that $\sigma$ goes to a constant, $\sigma=C$, at a large $\xi$,
which agrees with the general analysis presented in Paper I. As this solution
is obtained under the condition $\sigma\ll 1$, one concludes that there
should be $C\ll 1$. With account of
Eq. (\ref{small_sigma1}), Eq. (\ref{23}) yields
\eqb
\Gamma=1+C\left(\ln(1+\xi^2)-\frac{\xi^2}{1+\xi^2}\right).
 \label{small_sigma2}\eqe
Expanding Eqs. (\ref{small_sigma1}) and (\ref{small_sigma2}) in small $\xi$, one sees that the constant $C$
in this solution is the same that in Eq. (\ref{init1}).
Recall that the solution (\ref{small_sigma1}) and (\ref{small_sigma2}) was obtained under the assumption
$(\Gamma-1)^2\ll\sigma\ll 1$, which implies $C\ll 1$ and $\ln(1+\xi^2)\ll
1/\sqrt{C}$. The last condition shows that the solution (\ref{small_sigma1}) and (\ref{small_sigma2}) is valid from $\xi=0$ to
a large $\xi$ but not too large,
 \eqb
\xi\ll \exp\left(\frac 1{2\sqrt{C}}\right).
 \label{cond_xi1}\eqe
Let us now find the solution for an unrestrictedly large $\xi$, which
could be matched with the solution (\ref{small_sigma1})  and  (\ref{small_sigma2}).

One can easily find the solution to  Eqs. (\ref{22}) and (\ref{23}) at the condition
 \eqb
 \sigma\xi^2\gg\Gamma^2.
 \label{cond_xi2}\eqe
%Note that the regions (\ref{cond_xi1}) and (\ref{cond_xi2}) are intersected provided $C\ll 1$.
Then Eq. (\ref{equil1}) is reduced to
\eqb
\xi\frac{d\sigma}{d\xi}=\left(\frac{\Gamma}{\xi}\right)^2\ll \sigma;
 \eqe
which yields $\sigma=\it const$. This solution is matched with the solution
(\ref{small_sigma1}) if
$\sigma=C$ so that the solution (\ref{small_sigma1}) could be continued to an arbitrary large $\xi$.
In the same limit, Eq. (\ref{equil2}) is reduced to
  \eqb
\frac{\xi}{\Gamma}\frac{d\Gamma}{d\xi}=2\sigma=2C;
  \label{gamma}\eqe
which yields $\Gamma=C_1\xi^{2C}$, where $C_1$ is a constant.
In the region (\ref{cond_xi1}), this solution is reduced to
$\Gamma=C_1(1+2C\ln\xi)$, which is smoothly matched with the solution (\ref{small_sigma1}) provided $C_1=1-C$. Now one can write
 \eqb
\Gamma=(1-C)\xi^{2C}.
 \label{small_sigma4}\eqe
Recall that we solved the equations under the condition (\ref{cond_xi2}).
The function (\ref{small_sigma4}), together with the function $\sigma=C\ll 1$, satisfy this condition at $\xi\gg 1/\sqrt{C}$.
%Hence the function (\ref{small_sigma1}) provides the expression for $\sigma$ across the jet.
The Lorentz factor of the flow is given by the expression (\ref{small_sigma2}) at
$\ln\xi\ll 1/\sqrt{C}$ and by the expression (\ref{small_sigma4}) at $\xi\gg 1/\sqrt{C}$,
these two expressions
being smoothly matched in the region
\eqb
\frac 1{\sqrt{C}}\ll\xi\ll \exp\left(\frac 1{2\sqrt{C}}\right).
\eqe
One finally concludes that
Eqs. (\ref{small_sigma1}), (\ref{small_sigma2})  and (\ref{small_sigma4}) represent the
full solution for the transverse structure of a low-$\sigma$ jet.

The dependence of the jet structure on the distance from the origin, $Z$, enters via
the constant $C$, which is found from the boundary conditions (\ref{boundary3}) and (\ref{boundary2}). They could be written, substituting $C$ for $\sigma(\xi_0)$ and Eq. (\ref{small_sigma4}) for $\Gamma(\xi_0)$,  as
 \begin{eqnarray}
 && (1-C^2)\xi^{2C}_0=\frac{\gamma_{\rm max}}{\gamma_{\rm in}}; \\
&& \frac C{\xi_0}=\sqrt{\frac{\beta}3}\gamma_{\rm in}\gamma_{\rm max}Z^{\kappa/2}.
 \end{eqnarray}
Taking into account that $C\ll 1$, one gets
 \begin{eqnarray}
&& C=\frac{\ln(\gamma_{\rm max}/\gamma_{\rm in})}{2\ln\xi_0} \label{C}\\
&& \frac Z{Z_{\rm equipart}}=\left(\frac{2\gamma_{\rm in}\xi_0\ln\xi_0}
{\gamma_{\rm max}\ln(\gamma_{\rm max}/\gamma_{\rm in})}\right)^{2/\kappa}
 \label{shape}\end{eqnarray}
The last two equations provide dependence of the constant $C$ on $Z$ thus closing the solution
for the low-$\sigma$ part of the jet.

%Taking into account that $\gamma_{\rm max}\gg\gamma_{\rm in}$, one can approximately
%present the jet radius as
% \eqb
%X_0=%\frac{\ln(\gamma_{\rm max}/\gamma_{\rm in})}
%{\ln(\gamma_{\rm max}/\gamma_{\rm in})+(\kappa/2)\ln(Z/Z_{\rm tr})}
%\frac{\gamma_{\rm max}}{2}\left(\frac Z{Z_{\rm tr}}\right)^{\kappa/2}.
% \label{jet_rad}\eqe
Eq. (\ref{shape}) describes the shape of the flow lines at the $\sigma\ll 1$ stage, see Fig. 1.
Taking into account that this solution is valid at $\kappa<2$, one sees that the collimation angle continuously decreases, $d\xi_0/dZ\to 0$, so that the jet becomes asymptotically cylindrical.
According to Eq. (\ref{small_sigma1}), $C$ is equal to $\sigma$ at $\xi\gg 1$,
i.e. not too close to the axis.
Combining Eqs. (\ref{C}) and (\ref{shape}), one finds an estimate
 \eqb
\sigma(\xi\gg 1) = \frac 12\frac{\ln(\gamma_{\rm max}/\gamma_{\rm in})}
{\ln(\gamma_{\rm max}/\gamma_{\rm in})+(\kappa/2)\ln(Z/Z_{\rm equipart})}.
 \label{sigma_small}\eqe
Note that $\sigma$ is constant across the jet at $\xi\gg 1$; the full solution presented in Fig. 2 confirms this asymptotic result.
According to Eq. (\ref{sigma_small}), $\sigma$ becomes of the order of unity at $Z\sim Z_{\rm equipart}$,
so that the obtained asymptotics is roughly
matched with the asymptotics (\ref{gamma_sigma}-\ref{radius}) for the Poynting dominated
jet. An important point is that at $Z>Z_{\rm equipart}$,
$\sigma$ decreases however extremely slowly. For example, one sees in
Fig. 2 that for the chosen parameters of the jet, $\sigma$ decreases only to $1/3$ at the distance as large as $10^4Z_{\rm equipart}$.  The flow
becomes truly matter dominated, $\sigma<0.1$, only at
\eqb
Z>Z_{\rm equipart}\left(\gamma_{\rm max}/\gamma_{\rm in}\right)^{8/\kappa}.
\label{Zeq}\eqe

The results of this section are valid only for equilibrium jets; the corresponding condition
is given by Eq. (\ref{equilibrium_cond}). It follows from Eq. (\ref{radius}) that the Poynting
dominated jet is collimated in the equilibrium mode if the outer pressure decreases slowly
enough, $\kappa<2$. At the matter dominated stage, the flow is collimated slower. One sees
from Eq. (\ref{shape}) that only if $\kappa<1$, the flow at this stage satisfies the
condition (\ref{equilibrium_cond}) till the infinity. At $1<\kappa<2$, the flow expands faster
than $Z\propto X^2$ therefore eventually the condition
for the equilibrium collimation is violated. This occurs at the distance
 \eqb
Z=2^{\frac 2{\kappa-1}}\left(\frac{\beta}3\right)^{\frac 2{\kappa(\kappa-1)}}
\gamma_{\rm max}^{\frac{2(2-\kappa)}{\kappa(\kappa-1)}}.
 \eqe
The transition from the equilibrium to the non-equilibrium regime could be
studied only numerically.

\section{Non-equilibrium jets}

In this section, we study transition through $\sigma\sim 1$ in non-equilibrium jets.
The jet is efficiently accelerated in the non-equilibrium regime only if $\kappa=2$; $\beta<1/4$ (Paper I). At $\kappa<2$ the jet is accelerated in the equilibrium regime.
In the case
$\kappa=2$; $\beta>1/4$, the acceleration occurs in the intermediate regime when the jet is not in the transverse equilibrium however one cannot neglect the poloidal field; practically this intermediate regime is close to the equilibrium one.  At $\kappa>2$
the jet is in the non-equilibrium regime  however,
the acceleration is saturated at a terminal Lorentz factor,
which is generally less than $\gamma_{\rm max}$.

Of course if the initial Poynting flux is not too large, the flow could be accelerated till the $\sim\gamma_{\rm max}$
even if $\kappa>2$. For example, even though the non-confined flow is accelerated only till the terminal Lorentz factor $\gamma\sim(\gamma_{\rm max}\ln r)^{1/3}$ (Tomimatsu 1994; Beskin et al. 1998), the
equipartition is reached close enough to the axis, where the Poynting flux does not exceed this limiting value (Lyubarsky \& Eichler 2001; Tchekhovskoy, McKinney \& Narayan 2009a).
Another example is the flow confined by the external pressure with $\kappa$ something larger than 2. Then the flow is accelerated to the terminal Lorentz factor
\eqb
\gamma_t\sim (\gamma_{\rm max}/\Theta^2)^{1/3};
\label{gamma_t}\eqe
where $\Theta$ is the final collimation angle,
which is determined by the outer pressure distribution (Paper I). Such a flow could reach
equipartition only if $\gamma_{\rm max}\le\gamma_t$. An important point is that this is possible only if
$\gamma_{\rm max}\Theta\le 1$ so that such an efficient transformation of the Poynting to the kinetic energy anyway occurs only in causally connected, equilibrium flows. One should note that even if $\gamma_{\rm max}>\gamma_t$, so that the bulk of the flow stops accelerating still being Poynting dominated, the flow in the boundary layer is accelerated till the equipartition provided the confining pressure decreases; the width of this boundary layer is determined by the condition of the causal connection with the boundary  (Paper I, see also fig. 3a in Tchekhovskoy, Narayan \& McKinney 2009b).  We do not address all these cases in this paper and concentrate on the conditions permitting unrestricted acceleration up to $\gamma\sim\gamma_{\rm max}$.
The non-equilibrium jet could be accelerated till $\gamma\sim\gamma_{\rm max}$
 for any $\gamma_{\rm max}$ only if $k=2$, $\beta<1/4$. Here we study this case.

\begin{figure*}
\includegraphics[scale=0.8]{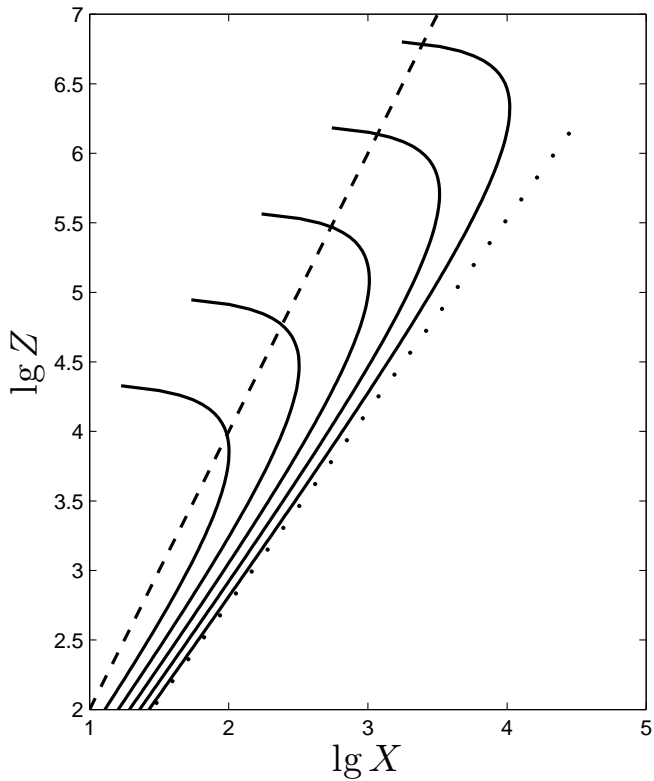}
\caption{The flux surfaces in the non-equilibrium jet; $\kappa=2$; $\beta=0.2$; $k=0.72$; $\gamma_{\rm max}=100$. Dotted line shows the Poynting dominated asymptotics $X=Z^k=Z^{0.72}$.
Dashed line shows the boundary $X=\sqrt{Z}$ within which the poloidal field could not be neglected
therefore the presented solution is valid only to the right of the dashed line.}
\end{figure*}

The non-equilibrium jet is described by the transfield equation in the form (\ref{nonequilibrium}), which should be supplemented by the Bernoulli equation (\ref{Bernoulli0}). Eliminating $\gamma$ from these equations and making use of Eq. (\ref{energy1})
for $\mu$, one gets a single equation, which looks in the dimensionless coordinates $X$ and $Z$ as
 \eqb
-\mu\frac{\partial^2X}{\partial Z^2}=\frac 2{2\mu\frac{\partial X}{\partial\mu}-X}
\frac{\partial}{\partial\mu}\frac X{2\mu\frac{\partial X}{\partial\mu}-X}.
 \label{noneq_tr}\eqe
This equation is invariant with respect to the transformation $\mu\to s\mu$,
$X\to s^mX$, $Z\to s^{m+1}Z$, where $s$ and $m$ are arbitrary numbers,
therefore one can look for a self-similar solution in the form
 \eqb
X=\mu^mU(\zeta);\quad \zeta=\frac Z{\mu^{m+1}}.
 \label{selfsim_variables}\eqe
With this ansatz, Eq. (\ref{noneq_tr}) is reduced to an ordinary differential equation for $U$:
 \begin{eqnarray}
&& \left[(2m-1)U-2(1+m)\zeta U'\right]^3U'' \nonumber\\
&& =4(1+m)^2\zeta [UU'+\zeta UU''-\zeta U'^2];
 \label{selfsim_eq}\end{eqnarray}
where the prime denotes derivative with respect to $\zeta$.

The general solution to Eq. (\ref{selfsim_eq}) is presented in Appendix.
We are interested in a solution satisfying
the boundary condition (\ref{boundary}) with the pressure distribution (\ref{pressure}), $\kappa=2$.
With the ansatz (\ref{selfsim_variables}), this boundary condition is written, with the aid of the Bernoulli equation (\ref{Bernoulli0}), as
 \eqb
(2m-1)U-2(m+1)\zeta U'=\sqrt{\frac 3{\beta}}\,\zeta.
 \label{boundary4}\eqe
Here we used Eq. (\ref{energy1}) at $\Psi=\Psi_0\gg\widetilde{\Psi}$. So we are looking for
a function satisfying both Eq. (\ref{selfsim_eq}) and Eq. (\ref{boundary4}). The solution to
Eq. (\ref{boundary4}) is written as
 \eqb
U=\frac 1{\sqrt{3\beta}}\left(\zeta_0^{1-k}\zeta^k-\zeta\right);
 \label{U_noneq}\eqe
where $\zeta_0$ is a constant,
 \eqb
k=\frac{2m-1}{2(1+m)}.
 \eqe
One can check straightforwardly that this function satisfies also Eq. (\ref{selfsim_eq}) provided
 \eqb
k=\frac 12\left(1+\sqrt{1-4\beta}\right).
 \eqe
Note that $1/2<k<1$. This solution exists only if $\beta<1/4$, which agrees with the conclusion in Paper I that at $\kappa=2$, the flow is in the non-equilibrium regime only at $\beta<1/4$.

Returning to the physical variables (\ref{selfsim_variables}), one
finds the shape of the flux surfaces (see Fig. 3)
 \eqb
X= \sqrt{\frac{\mu}{3\beta}}\,\zeta_0^{1-k}Z^k
\left[1-\frac 1{\mu^{3/2}}\left(\frac{Z}{\zeta_0}\right)^{1-k}\right].\label{Xnoneq}
 \eqe
The Bernoulli equation (\ref{Bernoulli0}) provides the expression for the Lorentz
factor of the flow
 \eqb
\gamma= \frac{3\mu\zeta^{1-k}}{\zeta_0^{1-k}+2\zeta^{1-k}}.\label{gamma_noneq}
 \eqe
At $\zeta\ll\zeta_0$, one comes to the scaling
 \eqb
 X\propto Z^k;\quad \gamma\propto Z^{1-k}
 \eqe
obtained earlier (Komissarov et al 2009; Paper I) for a Poynting dominated jet at $\kappa=2$, $\beta<1/4$. The expressions
(\ref{Xnoneq}) and (\ref{gamma_noneq}) generalize this solution beyond the Poynting dominated
domain. Note that when $\beta$ goes to $1/4$ from below, $k$ goes to $1/2$ so that
one comes to parabolic flow lines,
$X\propto\sqrt{Z}$, in which case one has to take into account the neglected poloidal field.
In the opposite limit, $\beta=0$,  one comes to an unconfined radial flow, $k=1$.

The constant $\zeta_0$ should be found by matching
this solution with the near zone solution. If the flow is not collimated at the light cylinder,
$X\sim 1$, $Z\sim 1$, there should be $\zeta_0\sim\gamma_{\rm max}^{-1/[2(1-k)]}$. Then one concludes that the jet ceases to be Poynting dominated  at the distance
 \eqb
Z_{\rm equipart}\sim\zeta_0\gamma_{\rm max}^{3/[2(1-k)]}\sim \gamma^{1/(1-k)}_{\rm max};
  \eqe
which agrees with the estimate presented in Paper I.

One sees that when $\zeta$ approaches $\zeta_0$, the flow converges to the axis and
the Poynting flux is converted into the kinetic energy, $\gamma\to\mu$. Note that at the Poynting dominated stage, $\zeta\ll\zeta_0$, the flow was causally disconnected,
\eqb
\frac{X\gamma}Z=\sqrt{\frac 3{\beta}}>1.
 \eqe
When it approaches equipartition, the bulk Lorentz factor does not grow any more, the causal connection is established and the magnetic loops squeeze the flow.
Recall that all
the results of the present section are obtained at the assumption that one can neglect
the poloidal field; the corresponding condition is given by Eq. (\ref{nonequilibrium_cond}).
If the flow remains axisymmetrical, it stops converging in the region $X\sim\sqrt{Z}$ where
the pressure of the poloidal field becomes significant.  However, one can expect that due to the kink instability, different magnetic loops could come apart in the converging flow forming an irregular field structure, which could trigger dissipation processes.

\section{Conclusions}

In this paper, we studied the acceleration of Poynting dominated
jets to the matter dominated
stage. Even though efficient transformation of the electromagnetic into the
kinetic energy is possible in principle in the scope of ideal MHD, the conditions for such a transition are not trivial. Namely,
the flow should be confined by external pressure, which decreases with the distance
but not too fast. It was shown in Paper I that the flow is accelerated until it ceases to be Poynting dominated only if $\kappa\le 2$ (for the power law confining pressure distribution (\ref{pressure})).
In the opposite case, the flow practically
stops accelerating after it reaches a terminal Lorentz factor of Eq. (\ref{gamma_t}), which is generally less than $\gamma_{\rm max}$. If $\gamma_t>\gamma_{\rm max}$, the flow behaves essentially as in the $\kappa<2$ case.

Especially restrictive is the fact that the acceleration zone spans a large range of scales so that one has to ensure that the conditions for acceleration
are fulfilled along all the way. If the confining pressure
begins to decrease faster than
$\propto z^{-2}$ or drops down abruptly, which happens for example when the GRB jet escapes from the progenitor star, the acceleration and collimation cease and the jet propagates further out preserving the acquired collimation angle and the Lorentz factor.
%\footnote{Efficient acceleration found by Tchekhovskoy, McKinney \& Narayan (2009b) in their simulations of %the jet escaping from the progenitor star should be attributed to the boundary effect. At the boundary of %the flow, the Poynting flux is in any case converted into the kinetic energy  provided the external %pressure decreases however, the width of the accelerated layer is determined by the causality (Paper I). %Since $\gamma\Theta\gg 1$ in GRB jets, the width of the efficiently accelerated layer should be small.  %Tchekhovskoy et al simulated a jet with a constant angular velocity, in which most of the Poynting flux is %concentrated at the periphery, therefore the boundary effects were important in their case. Real jets are %presumably isolated from the boundary by the wind from the accretion disc. In outflows with the angular %velocity decreasing outward, the central part, where most of the energy is transferred, is not accelerated %if the pressure of the confining medium sharply drops down. }. 
On the other hand, if the confining pressure stops decreasing, the flow becomes cylindrical however the acceleration is terminated.

The flow is efficiently accelerated in the equilibrium regime, i.e. if at any distance from the origin, the structure of the flow is settled into the structure of an appropriate cylindrical configuration. At the Poynting dominated stage, the equilibrium flow is accelerated as
$\gamma\sim \Omega r$ (Tchekhovskoy et al 2008; Paper I, Beskin \& Nokhrina 2009) therefore
the faster the flow expands, the faster it is accelerated. On the other hand, the condition
(\ref{equilibrium_cond}) implies that the flow is in the equilibrium only if it remains
within the parabola $X\sim\sqrt{Z}$ (which in fact ensures that the causal connection is maintained accross the flow) therefore the acceleration rate is maximal for the parabolic flow.
The flow expansion is determined by the distribution of the confining pressure. The faster the outer pressure decreases,
the faster the jet expands. Taking into account Eqs. (\ref{equilibrium_cond}) and (\ref{radius}), one sees that at $\kappa<2$ the flow is in the equilibrium regime\footnote{If $\kappa$ is a bit larger than 2 and $\beta>1$, the equilibrium conditions still could be met not too far from the source. In this limited region, the flow behaves as an equilibrium one, i.e. it expands and accelerates according to the equilibrium scalings (\ref{gamma_eq}) and (\ref{radius}), see paper I.}.

The fastest acceleration regime %, $\gamma=X=(3/\beta)^{1/4}\sqrt{Z}$,
is achieved when $\kappa$ goes to 2.
At $\kappa=2$, the flow is in the equilibrium regime only at $\beta\gg 1$. The case $\beta\sim 1$ is an intermediate between the equilibrium and non-equilibrium regimes. At the Poynting dominated stage, the properties of the $\kappa=2$ flow are very similar to those of the equilibrium flow at $\beta>1/4$ (Komissarov et al. 2009, Paper I). In this case,
the jet has a parabolic form
 \eqb
X=\left(\frac 3{\beta-1/4}\right)^{1/4}Z^{1/2};
 \eqe
whereas the Lorentz factor grows as $\gamma=X$.
The energy equipartition is achieved in such a flow at the distance %(cf. Eq. (\ref{Ztr}))
$Z_{\rm equipart}\sim\gamma_{\rm max}^2$.
For a slower decreasing external pressure, $\kappa<2$,
the flow remains Poynting dominated at even larger distances, see Eq. (\ref{Ztr}).

GRB jets are known to have Lorentz factors of at least a few hundreds (Lithwick \& Sari 2001).
In the fastest acceleration regime, one could reach these Lorentz factors if the size of the acceleration region is $\sim 10^5(\gamma/300)^2$ of the light cylinder radii. Taking into account that the characteristic light cylinder radius of a rapidly rotating black hole of a few solar masses is $\sim 10^6\div 10^7$ cm, one sees that the above estimate of the acceleration zone is compatible with the size of the progenitor star. The problem is that the observations of the afterglow light curves as well as the burst statistics evidence for $\gamma\theta\sim 10\div 30$ (e.g. Tchekhovskoy et al, 2009b) whereas an efficient transformation of the Poynting to the kinetic energy occurs only if  $\gamma\theta\sim 1$. Highly collimated but causally disconnected jets could be formed if the confining pressure decreases something faster than $z^{-2}$ (Paper I) however, such jets remain Poynting dominated so that magnetic dissipation is necessary in order to utilize the energy of the outflow.

Observations of jets in AGNs and microquasars evidence for Lorentz factors from a few to a few dozens (e.g. Cohen et al 2003; Mirabel \& Rodri\'guez 1999). In order to achieve these Lorentz factors in the fastest acceleration regime, one needs the size of the confinement zone of only $Z_{\rm equipart}\sim 100\div 1000r_g$ provided the black hole is rapidly rotating. The wind from the accretion disk could serve as the confining medium up to distances of the order of the external disk radius, which could be that large. It is not clear whether the wind from the disk could provide the confining pressure decreasing not faster than $r^{-2}$. If the pressure decreases faster, the jet remains Poynting dominated. An important point is that even if the necessary conditions are fulfilled, the flow could reach only an equipartition state but not a true matter dominated stage, $\sigma<0.1$.
In this paper we have shown that when the equilibrium jet ceases to be Poynting dominated,
the collimation angle still decreases even though slower than at the Poynting dominated stage.
However, $\sigma$ decreases only logarithmically so that $\sigma\approx 0.1$ is achieved only if the confining medium is extended beyond the distance  $Z_{\rm equipart}\gamma^4_{\rm max}
\sim 10^6\div 10^7(\gamma_{\rm max}/10)^4r_g$ (see Eq. (\ref{Zeq}), we assumed that $\gamma_{in}\sim 1$),
which seems to be inappropriately large. The fact that without magnetic dissipation, jets could not become true matter dominated, has important implications for the interaction of the ejected material with the surroundings (Leismann et al 2005; Mimica \& Aloy 2009; Mimica et al. 2009).

Since the causal contact is maintained across equilibrium jets, one has to worry about the kink instability, which could significantly disturb or even destroy the regular flow structure. However, last studies reveal (Tschechovskoy et al. 2008) that in Poynting dominated outflows,
the poloidal field is very close to uniform (and is exactly uniform for the chosen here simple expression (\ref{energy1}) for $\mu(\Psi)$ and $\Omega=\it const$, $\eta=\it const$); in this case the growth rate of the kink instability goes to zero  (Istomin \& Pariev 1996, Lyubarskii 1999). It is possible that the instability could develop in spite of the low growth rate because the jet acceleration zone is very large but in order to clarify the question, more careful investigation of the transverse structure of the jet is necessary. Note also that even if the instability turns out to be suppressed in the Poynting dominated stage, it could develop at the moderately magnetized stage when the poloidal field is concentrated to the axis of the flow (Paper I, Beskin \& Nokhrina 2009). In any case the impact of the instability on the jet structure should be studied only with 3D numerical simulations.

The acceleration in the non-equilibrium regime is generally not very efficient
so that the flow Lorentz factor could not significantly exceed a terminal value
determined by the
parameters of the flow and of the surrounding medium (Paper I). Only in the specific case $\kappa=2$, $\beta<1/4$,  the jet is accelerated in non-equilibrium regime till the equipartition level.
We showed that after such a flow reaches rough equipartition, it sharply converges to the axis  and
the energy is efficiently transferred to the plasma. It is not clear what happens
to this "collapsing" flow; one can expect that a sort of a "hot spot"
appear in such a flow, which resembles that formed by hydrodynamical recollimation of a relativistic outflow (Levinson \& Bromberg 2008; Bromberg \& Levinson 2009).

In any case, we have not found, in the scope of ideal MHD,
a possibility for a smooth acceleration of a Poynting
dominated flow to the matter dominated stage at a reasonable scale. In equilibrium jets,
$\sigma$ decreases too slowly after the flow ceases to be Poynting dominated.
In the only case of efficiently
accelerated non-equilibrium jet, the flow is not continued till the infinity but in fact
collapses.

In this paper, we have addressed the transition to the matter dominated stage
both in the equilibrium and in the non-equilibrium flows. The intermediate case,
$\kappa=2$; $\beta\approx 1$, is beyond the scope of our analytic approach.
At the Poynting dominated stage, the behavior of the intermediate flow is
similar to the behavior of the equilibrium flows (Komissarov et al 2009; Paper I).
In order to find what happens to such a flow at the $\sigma<1$ stage, one has to solve Eqs. (\ref{asympt_transfield}) and (\ref{Bernoulli0}) numerically.

\section*{Acknowledgments}
The work is supported by the US-Israeli Binational Science foundation (grant 2006170) and by the Israeli Science Foundation (grant 737/07).

\appendix
\section{}%General solution to Eq. (\ref{selfsim_eq})}
The general solution to Eq. (\ref{selfsim_eq}) is found taking into account that
this equation is invariant under the transformation $\zeta\to s\zeta$, $U\to sU$.
Therefore we can reduce the order of the equation using the substitution
 \eqb
f=U/\zeta;\quad \phi=-\frac{\zeta}f\frac{df}{d\zeta}.
 \label{variable}\eqe
Then one gets a linear first order equation,
\eqb
\frac 12(1-\phi)\frac{df^2}{d\phi}=f^2+\frac{4(1+m)^2}{[3-2(1+m)\phi]^3};
 \eqe
which is immediately solved to yield
 \eqb
f^2(\phi)=\frac{3k}{1-k}\left(\frac 1{(\phi-k)^2}-\frac{c^2}{(1-\phi)^2}\right);
 \eqe
where  $c$ is a constant.
Substituting then $f(\phi)$ to the second equation (\ref{variable}), one gets
a differential equation
for $\zeta(\phi)$, which is easily solved.

Finally one gets the solution to
Eq. (\ref{selfsim_eq}) in the parametric form:
\begin{eqnarray}
&& U(\phi)=\zeta_0\sqrt{\frac{(1-k)(1-c^2)}{3k}}\phi^{-\frac{1-(1-k)^3c^2}{(1-k)[1-(1-k)^2c^2]}}
\label{U}\\
&&\times\vert\phi-1+k\vert^{k/(1-k)}\vert k_1-\phi\vert^{-(1-k_1)/(2k_1)}\vert k_2-\phi\vert^{-(1-k_2)/(2k_2)};\nonumber\\
&& \zeta(\phi)=\zeta_0|1-\phi |\phi^{-\frac{1-(1-k)^3c^2}{(1-k)[1-(1-k)^2c^2]}}\label{zeta}\\
&&\times\vert\phi-1+k\vert^{1/(1-k)}\vert k_1-\phi\vert^{-1/(2k_1)}\vert k_2-\phi\vert^{-1/(2k_2)}\nonumber
 \end{eqnarray}
where  $\zeta_0$ is a constant,
 \eqb
k_{1,2}=1\mp\frac{kc}{1\pm c}.
 \eqe
Substituting the expressions (\ref{U}) and (\ref{zeta}) into Eq. (\ref{selfsim_variables}),
one gets a set of non-equilibrium jet configurations depending on three parameters, $m$ (or $k$), $c$ and $\zeta_0$. The Bernoulli equation (\ref{Bernoulli0}) provides
the expression for the Lorentz factor of the flow:
 \eqb
\gamma=\mu\frac{\phi-1+k}{\phi-(2/3)(1-k)}.
 \label{gamma1}\eqe
One sees that there should be $\phi>1-k$ in order for $\gamma$ to be positive.
Note also that $\gamma\to \mu$ (i.e. $\sigma\to 0$) only if $\phi\to\infty$.
The solution (\ref{U_noneq}), (\ref{gamma_noneq}) is obtained from the general
solution (\ref{U}), (\ref{zeta}) and (\ref{gamma1}) at $c=0$.

\end{document}